\documentstyle[sprocl,epsfig,color]{article}

\def\be{\begin{equation}}
\def\ee{\end{equation}}
\def\bea{\begin{eqnarray}}
\def\eea{\end{eqnarray}}

\def\d{\mathrm{d}}

\def\be#1\ee{\begin{equation}#1\end{equation}}
\def\ba#1\ea{\begin{align}#1\end{align}}

\def\aitalc{{\sc \textit{a}{\r{\i}}\raisebox{-0.14em}{T}alc}}

\newcommand{\diana}{{\sc{Diana}}}
\newcommand{\qgraf}{\textsc{Qgraf}}
\newcommand{\form}{\textsc{Form}}
\newcommand{\fortran}{\textsc{Fortran}}

\newcommand{\looptools}{\textsc{LoopTools}}

\newcommand{\tab}[1]{Tab.~\ref{#1}}

\renewcommand{\d}[1]{\mathrm{d}#1}

\def \Oa{\mathcal{O}(\alpha)}

\def \ct{\cos{\theta}}
\def \emo{\cdot{} 10^{-1}}

\begin{document}
\begin{flushleft}
{\normalsize \tt 
SFB/CPP-04-38
\\ 
DESY 04-161
\\ March 2004}
\end{flushleft}
\vspace*{.2cm}

\title{AUTOMATIZED CALCULATION OF 2-FERMION PRODUCTION WITH \protect
  \diana{} AND \protect \aitalc{} 
 \footnote{Work supported in part 
  by European's 5-th Framework under contract HPRN--CT--2000--00149 Physics at
  Colliders and
  by Deutsche Forschungsgemeinschaft under contract SFB/TR 9--03.}\\}

\author{J. FLEISCHER, A. LORCA, T. RIEMANN}
\vspace{.4cm}
\address{Deutsches Elektronen-Synchrotron, DESY,
   Platanenallee 6, 15738 Zeuthen, Germany}


\maketitle

\abstracts%
{%
The family of two fermion final states is among the cleanest final
states at the International Linear Collider (ILC) project.
The package {\aitalc} has been developed for a calculation of their production
cross sections, and we present here 
benchmark numerical results in one loop approximation in the
electroweak Standard Model.
We are using packages like \qgraf, \diana, \form, \looptools{} for intermediate
steps.}
\section{Introduction}
The family of two fermion final states is among the cleanest final
states at the International Linear Collider (ILC) project
\cite{Aguilar-Saavedra:2001rg}: 
\bea
\label{1}
e^+e^- &\to& t {\bar t} (\gamma),
\\
\label{2}
e^+e^- &\to& b {\bar b} (\gamma), \hspace{0.3cm} \mu {\bar \mu}
  (\gamma), \ldots,
\\
\label{3}
e^+e^- &\to& e^+ e^- (\gamma),
\\
\label{4}
e^+e^- &\to& s {\bar b},  \hspace{0.3cm}  c {\bar t},
  \hspace{0.3cm}  \mu {\bar \tau}, \ldots 
\eea
Some of these reactions have high statistics in the Standard Model (SM)
and allow high precision measurements, others are forbidden or rare in
the SM and might deliver signals of New Physics.
Reaction (\ref{1}) was studied in
\cite{Fleischer:2003kk,Biernacik:2003xv}, 
reactions  (\ref{2}) in \cite{Hahn:2003ab}, 
reaction (\ref{3}) in \cite{Gluza:2003nn2,Lorca:2004dk2},
and (\ref{4}) in \cite{Lorca:2004dk2}, and in references therein.  
\section{\protect \diana{} and \protect \aitalc}
\begin{figure*}[t]
\begin{center}
\includegraphics[width=11.cm]{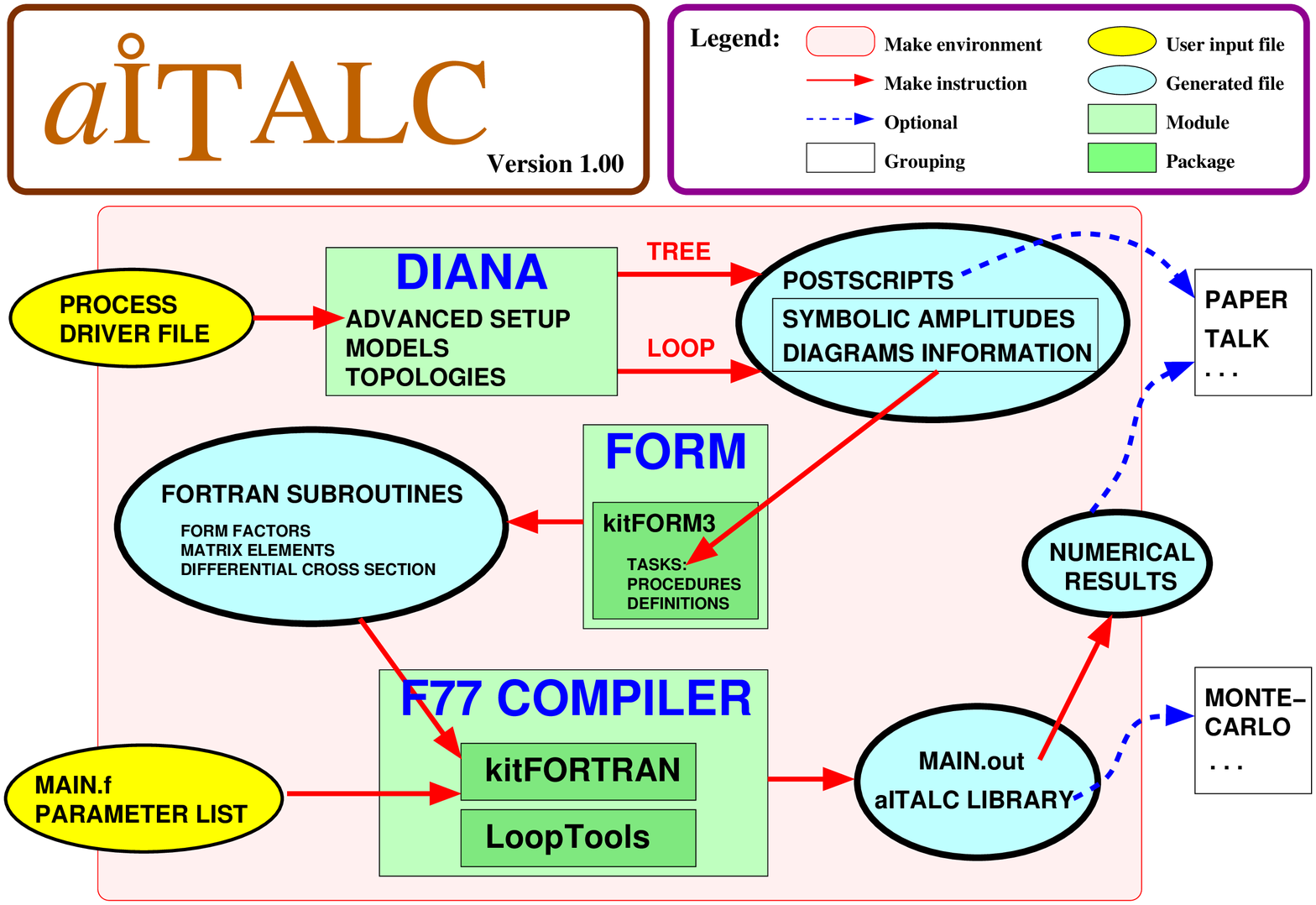}
\vspace{-7pt}
\caption{Flow chart of the \protect \aitalc{} package.}
\label{flowchart}
\end{center}
\end{figure*}

\diana{} \cite{Tentyukov:2002ig} is a package for the analysis of Feynman diagrams and their
further calculation with \form{} \cite{Vermaseren:2000nd}, but has also some useful
graphical output. 
It integrates \qgraf{} \cite{Nogueira:1993ex}, a package for the creation of all
Feynman diagrams for a given reaction in a given model.
With \aitalc{} we go a step further and automatize the complete
calculation, including renormalization and the creation and running of
a \fortran{} code. 
For the final numerics, \looptools{} is used~\cite{Hahn:1998yk,vanOldenborgh:1991yc}. 
The symbolic structure of \aitalc{} is shown in Figure \ref{flowchart}.
The package with examples is available from \cite{Zeuthen-aITALC:2004a}.

\section{Numerical Results}
We use the same input values as in
\cite{Hahn:2003ab,Lorca:2004dk2}.  
They are described in Table \ref{IPS}.

The Tables \ref{table_tau_500} to  5 
contain sample
values of differential cross sections  at $\sqrt s = 500$
GeV for $e^+e^-\to  \tau {\bar \tau},~ b {\bar b}, ~ c {\bar
  c}, ~ e^+e^-$, 
respectively.  
The columns contain 
Born cross-sections and those including also the weak and soft
photonic $\Oa$ corrections.  

For the Bhabha case, we have extended the angular range in the forward
direction and also show QED corrections in order to illustrate
the dominance of pure photonic $t$ channel exchange diagrams.
The weak
corrections stay small or even negligible there at both GigaZ and ILC
energies. 
Bhabha forward scattering at 26 \dots 82 mrad is planned to be used
for the luminosity measurement at the $10^{-4}$ 
level \cite{Lohmann:2004nn,Lohmann:2004nn2}. 
This corresponds to $\cos \vartheta = 0.99966 \dots 0.99664$ and $|t| >$
1.36 GeV$^2$ (42.2 GeV$^2$) at GigaZ (ILC).
Since complete 2-loop contributions have to be taken into account for
this measurement,
the restriction to pure QED is an enormous simplification, although
the resulting task stays extremely nontrivial, nevertheless
\cite{Czakon:2004nn2}.

\begin{table}[htb]
\begin{center}
\begin{tabular}{|c|c|}
\hline
Fermion Masses & Boson Masses \& Widths\\
\hline
\begin{math}
\begin{array}{l@{=~}r@{.}l@{\mathrm{~GeV}}}
m_{\nu}&0&0\\
m_e&0&00051099907\\
m_{\mu}&0&105658389\\
m_{\tau}&1&77705\\
m_u&0&062\\
m_c&1&5\\
m_t&173&8\\
m_d&0&083\\
m_s&0&215\\
m_b&4&7
\end{array}
\end{math}
&
\begin{math}
\begin{array}{l@{=~}r@{.}l@{\mathrm{~GeV}}}
m_{\gamma}&0&0\\
m_W&80&4514958\\
m_Z&91&1867\\
m_H&120&0\\
\Gamma_W&0&0\\
\Gamma_Z&0&0\\
\Gamma_H&0&0
\end{array}
\end{math}
\\
\hline
\hline
\multicolumn{2}{|c|}{Other Parameters}\\
\hline
\multicolumn{2}{|c|}{
\begin{math}
\begin{array}{l@{=~}l}
\alpha&1/137.03599976\\
E^{\textrm{max}}_{\gamma_\textrm{soft}}& \phantom{\Big|}\sqrt{s}/10\\
(\hbar c)^2& 0.38937966 \cdot 10^9 ~\mathrm{GeV}^2 \textrm{pb}\\
\end{array}
\end{math}
}\\
\hline
\end{tabular}
\caption{Input parameter set.}
\label{IPS}
\end{center}
\end{table}

\def\begintab#1#2{%
  \begin{array}{|r|l|l|}
  \hline
  \multicolumn{3}{|c|}{\vphantom{\Big|}e^+e^-\to #1\qquad \sqrt s = \textrm{#2}} \\
  \hline\vphantom{\Big|}
  \ct &
  \left[\frac{\d\sigma}{\d\ct}\right]_{\textrm{Born}}/\textrm{pb} &
  \left[\frac{\d\sigma}{\d\ct}\right]_{\textrm{B+w+QED+soft}}/\textrm{pb} 
\\
  \hline\hline
}

\begin{table}[htb]
{
$$
\begintab{\tau^+\tau^-}{500 GeV}
-0.9 & 0.94591~02171~8632{9} \emo & 0.92419~02671~1{4061} \emo \\
-0.5 & 0.89298~53117~7985{8} \emo & 0.86699~48248~6{5248} \emo \\
 0.0 & 0.15032~16827~75192 & 0.14359~79492~086{48} \\
 0.5 & 0.28649~90174~53525 & 0.28258~86777~59{811} \\
 0.9 & 0.44955~18970~14604 & 0.47648~29191~20{038} \\
\hline
\end{array}
$$
}
\caption{Differential cross-sections for 
 $\tau$-production at $\sqrt s = 500$ GeV.  }
\label{table_tau_500}
\end{table}

\begin{table}[t]
{
$$
\begintab{b\bar b}{500 GeV}
-0.9 & 0.35947~21020~03927 \emo & 0.37629~38061~{ 51582} \emo \\
-0.5 & 0.52846~99142~9459{ 5} \emo & 0.49542~16119~6{ 4096} \emo \\
 0.0 & 0.13444~84372~56821 & 0.12117~62087~0{ 2347} \\
 0.5 & 0.28324~62378~51991 & 0.26454~12363~9{ 5596} \\
 0.9 & 0.45066~58537~60950 & 0.44708~31668~1{ 9343} \\
\hline
\end{array}
$$
}
\caption{The same as \tab{table_tau_500} for $b$-pair production at $\sqrt s = 500$ GeV.}
\label{table_b_500}
\end{table}

\begin{table}[t]
{
$$
\begintab{c\bar c}{500 GeV}
-0.9 & 0.78403~69156~96992 \emo & 0.83668~39315~{90920} \emo \\
-0.5 & 0.10411~12875~82399      & 0.10590~20427~16{561} \\
 0.0 & 0.24770~82888~4590{1}    & 0.23448~15990~2{5778} \\
 0.5 & 0.51515~25192~73431      & 0.46371~41775~1{7198} \\
 0.9 & 0.81827~79086~1355{7}    & 0.70026~97050~2{9472} \\
\hline
\end{array}
$$
}
\caption{The same as \tab{table_tau_500} for $c$-pair production at $\sqrt s = 500$ GeV.}
\label{table_c_500}
\end{table}

\def\begintabe#1#2{%
  \begin{array}{|r|l|l|c|}
  \hline
  \multicolumn{4}{|c|}{\vphantom{\Big|}e^+e^-\to #1\qquad \sqrt s = \textrm{#2}} \\
  \hline\vphantom{\Big|}
  \ct &
  \left[\frac{\d\sigma}{\d\ct}\right]_{\textrm{Born}}/\textrm{pb} &
  \left[\frac{\d\sigma}{\d\ct}\right]_{\textrm{B+1-loop}}/\textrm{pb} 
&\textrm{Model} 
\\
  \hline\hline
}
\begin{table}[t]
{
$$
\begintabe{e^+e^-}{500 GeV}
-.9000 & 0.216~998           & 0.144~359           & \mathrm{EWSM}\\
-.9000 & 0.523~873           & 0.387~798           & \mathrm{QED}\\
\hline                      
-.5000 & 0.261~360           & 0.181~086           & \mathrm{EWSM}\\
-.5000 & 0.611~600           & 0.471~451           & \mathrm{QED}\\
\hline
0.0000 & 0.598~142           & 0.431~573           & \mathrm{EWSM}\\
0.0000 & 0.117~253\cdot 10^1 & 0.916~946           & \mathrm{QED}\\
\hline
0.5000 & 0.421~272\cdot 10^1 & 0.320~045\cdot 10^1 & \mathrm{EWSM}\\
0.5000 & 0.550~440\cdot 10^1 & 0.435~535\cdot 10^1 & \mathrm{QED}\\
\hline
0.9000 & 0.189~160\cdot 10^3 & 0.150~885\cdot 10^3 & \mathrm{EWSM}\\
0.9000 & 0.189~118\cdot 10^3 & 0.152~861\cdot 10^3 & \mathrm{QED}\\
\hline
0.9900 & 0.206~555\cdot 10^5 & 0.170~576\cdot 10^5 & \mathrm{EWSM}\\
0.9900 & 0.206~381\cdot 10^5 & 0.170~818\cdot 10^5 & \mathrm{QED}\\
\hline
0.9990 & 0.208~236\cdot 10^7 & 0.176~139\cdot 10^7 & \mathrm{EWSM}\\
0.9990 & 0.208~242\cdot 10^7 & 0.176~190\cdot 10^7 & \mathrm{QED}\\
\hline
0.9999 & 0.208~429\cdot 10^9 & 0.180~172\cdot 10^9 & \mathrm{EWSM}\\
0.9999 & 0.208~430\cdot 10^9 & 0.180~178\cdot 10^9 & \mathrm{QED}\\
\hline
\end{array}
$$
}
\caption{%
Differential cross-sections for Bhabha scattering at $\sqrt s = 500$
GeV.
The running of $\alpha_{em}$ has been switched off in this table. 
For each angle, in the second line the weak corrections are switched
off in addition. 
}
\label{tab_bha_500}
\end{table}

\section*{Acknowledgement}
One of us (J.F.) would like to thank DESY for support. 

\section*{References}

\begingroup\begin{thebibliography}{10}

\bibitem{Aguilar-Saavedra:2001rg}
{ECFA/DESY LC Physics Working Group} Collaboration, J.~Aguilar-Saavedra {\em et
  al.},
\href{http://www.arXiv.org/abs/hep-ph/0106315}{{\tt hep-ph/0106315}}.

\bibitem{Fleischer:2003kk}
J.~Fleischer, A.~Leike, T.~Riemann, and A.~Werthenbach, {\em Eur. Phys. J.}
  {\bf C31} (2003) 37,
\href{http://www.arXiv.org/abs/hep-ph/0302259}{{\tt hep-ph/0302259}}.

\bibitem{Biernacik:2003xv}
A.~Biernacik, K.~Ko{\l}odziej, A.~Lorca, and T.~Riemann, {\em Acta Phys.
  Polon.} {\bf B34} (2003) 5487,
\href{http://www.arXiv.org/abs/hep-ph/0311097}{{\tt hep-ph/0311097}}.

\bibitem{Hahn:2003ab}
T.~Hahn, W.~Hollik, A.~Lorca, T.~Riemann, and A.~Werthenbach, LC-TH-2003-083,
\href{http://www.arXiv.org/abs/hep-ph/0307132}{{\tt hep-ph/0307132}}.

\bibitem{Gluza:2003nn2}
J.~Gluza, A.~Lorca, and T.~Riemann, {\em Automated use of DIANA for two-fermion
  production at colliders}, presented at ACAT 2003, KEK, Japan, 1-5 Dec 2003, {\em
  Nucl. Instrum. Meth.} {\bf A}, in press,
\href{http://www.arXiv.org/abs/hep-ph/0409011}{{\tt hep-ph/0409011}}.

\bibitem{Lorca:2004dk2}
A.~Lorca and T.~Riemann, to appear in {\em Nucl. Phys. (Proc. Suppl.)} {\bf B},
\href{http://www.arXiv.org/abs/hep-ph/0407149}{{\tt hep-ph/0407149}}.

\bibitem{Tentyukov:2002ig}
M.~Tentyukov and J.~Fleischer, {\em Nucl. Instrum. Meth.} {\bf A502} (2003)
570.

\bibitem{Vermaseren:2000nd}
J.~A.~M. Vermaseren,
\href{http://www.arXiv.org/abs/math-ph/0010025}{{\tt math-ph/0010025}}.

\bibitem{Nogueira:1993ex}
P.~Nogueira, {\em J. Comput. Phys.} {\bf 105} (1993) 279.

\bibitem{Hahn:1998yk}
T.~Hahn and M.~P{\'e}rez-Victoria, {\em Comput. Phys. Commun.} {\bf 118} (1999)
  153,
\href{http://arXiv.org/abs/hep-ph/9807565}{{\tt hep-ph/9807565}}.

\bibitem{vanOldenborgh:1991yc}
G.~J. van Oldenborgh, {\em Comput. Phys. Commun.} {\bf 66} (1991)
1.

\bibitem{Zeuthen-aITALC:2004a}
A. Lorca and T. Riemann, see webpage \\
  http://www-zeuthen.desy.de/theory/research/CAS.html.

\bibitem{Lohmann:2004nn}
W.~Lohmann, Contrib. to 2003 IEEE Nucl. Sci. Symp. and Med. Imag. Conf.,
  Portland, USA, to appear in the Proc.

\bibitem{Lohmann:2004nn2}
W.~Lohmann {\em et al.}, http://www-zeuthen.desy.de/lcdet/.

\bibitem{Czakon:2004nn2}
M.~Czakon, J.~Gluza, and T.~Riemann, to appear in {\em Nucl. Phys. (Proc.
  Suppl.) } {\bf B},
\href{http://www.arXiv.org/abs/hep-ph/0406203}{{\tt hep-ph/0406203}}.

\end{thebibliography}\endgroup
\providecommand{\href}[2]{#2}
\end{document}